\documentclass[11pt]{article}
\usepackage{amsmath,amsfonts,amsbsy}
\usepackage[dvips]{graphicx,epsfig}

\newcommand{\hoch}[1]{$^{#1}$}

\def\mso{\mathfrak{so}}

\def\bec{\begin{center}}
\def\ec{\end{center}}

\def\d{\delta} 

\def\e{\epsilon}

\def\be{\begin{equation}}
\def\ee{\end{equation}}
\def\bea{\begin{eqnarray}}
\def\eea{\end{eqnarray}}
\def\ba{\begin{array}}
\def\ea{\end{array}}

\begin{document}

\hfill{\texttt{SPIN-07/01}}

\vspace{-5pt}

\hfill{\texttt{ITP-UU-07/01}}

\vspace{-5pt}

\hfill{\texttt{KUL-TF-07/01}}

\vspace{-5pt}
\hfill{\texttt{hep-th/0701081}}

\hfill{\today}

\vspace{20pt}

\begin{center}

%%%%%%%%%%%%%%%%%%%%%%%%%%%%%%%%%%%%%%%%%%%%%%%%%%%%%%%%%%%%%%%%%%%%

{\Large\sc Singleton Strings}

%%%%%%%%%%%%%%%%%%%%%%%%%%%%%%%%%%%%%%%%%%%%%%%%%%%%%%%%%%%%%%%%%%%%

\vspace{30pt}
{\sc J. Engquist\hoch1, P. Sundell\hoch2 and L. Tamassia\hoch3}\\[15pt]

\hoch{1}{\it\small Institute for Theoretical Physics and Spinoza
Institute \\ Utrecht University \\3508 TD Utrecht, The
Netherlands}\vspace{5pt}

\vspace {0.5cm}

\hoch{2}{\it\small Scuola Normale Superiore and INFN\\ Piazza dei
Cavalieri 7, 56126 Pisa, Italy}\vspace{5pt}

\vspace {0.5cm}

\hoch{3}{\it\small Instituut voor Theoretische Fysica
\\Katholieke Universiteit Leuven \\
Celestijnenlaan 200D, B-3001 Leuven, Belgium}\vspace{30pt}

%%%%%%%%%%%%%%%%%%%%%%%%%%%%%%%%%%%%%%%%%%%%%%%%%%%%%%%%%%%%%%%%%%%%

{\sc\large Abstract}
\end{center}

In this note, we briefly introduce singleton representations and discuss their relevance to the study of string and brane configurations in AdS space, their tensionless limit and the connection with higher-spin gauge theory.
We then discuss the main properties of subcritical $SO(D-1,2)$ WZW models featuring singletons and their composites as primary fields. After a suitable gauging, these models are good candidates for a fundamental, partonic description of tensionless branes in AdS.
Their massless sector provides an affine Lie-algebraic setting for the study of higher-spin symmetries.
Based on hep-th/0508124, hep-th/0701051 and work in progress. 

\noindent 

{\vfill\leftline{}\vfill \vskip  10pt \footnoterule {\footnotesize
 This is the transcript of a talk given by L. T. at the RTN project `Constituents, fundamental forces and symmetries of the universe' conference in Napoli, October 9 -13, 2006.\vskip -12pt}}

\setcounter{page}{1}

\pagebreak

The group theoretical structure underlying physics in anti de Sitter
(AdS) spacetime is intrinsically different with respect to the flat
case, due to the presence of special ultra-short representations,
named singletons, that do not admit a flat space limit. The purpose
of this collaboration is to exploit 
this feature in the study of string and brane dynamics in AdS
spacetime, in particular while trying to establish a connection
between String Theory in AdS backgrounds (in the tensionless
limit) and Higher-Spin Gauge Theory (HSGT). This connection could
provide new insight into string quantization in AdS,
as well as more general backgrounds. As a byproduct, in \cite{ourpaper}, we have
exploited the gained physical intuition to study a class of
noncompact WZW models that are not yet well understood and
interesting in their own right.

Singleton representations play a key role in
the study of the AdS/CFT correspondence. On the CFT side, a partonic
description in terms of a discrete
spin chain has been proven to be very powerful \cite{spinchain}. In
this formulation, a singleton resides
at each site of the spin chain \cite{singlspinchain}.
A natural question is then whether a
corresponding microscopic description emerges on the string side in
an appropriate regime.\vskip 10pt

Singletons are ultra-short unitary
irreducible representations (UIR) of $\mso(D-1,2)$ whose
weights form single lines in weight space. This phenomenon was first
discovered by Dirac \cite{Dirac} in 1963 in the case of the scalar
and spinor singletons in $D=4$. Since $SO(D-1,2)$ can be realized as
the group of the isometries of AdS$_D$, singletons play a key role
in the study of AdS physics. As already mentioned, singletons do not
admit a flat space limit. The other unitary irreducible
representations of $\mso(D-1,2)$ can be classified as massless or
massive, according to their flat space limit. \\ The singletons also
play an important role in HSGT, where the presently known full
higher-spin field equations, due to Vasiliev \cite{Vasse}, are based
on gauging higher-spin algebras given by subalgebras of the enveloping algebra of $\mso(D-1,2)$ obtained by factoring out ideals given by singleton
annihilators.

The main characteristic of singletonic particles is that they are
unobservable in the bulk of AdS and, as isolated objects, they can
only be observed on the boundary. They can be naturally described by
a conformal particle on the zero radius limit of AdS, known as
Dirac's Hypercone, leading to an $Sp(2)$-gauged sigma model
\cite{zeror,bars}. However, their composites are indeed observable in the
bulk, since they are the ordinary massless and massive particles in
AdS.

In fact, a fundamental result for quantum field theory in AdS spaces
is the compositeness theorem by Flato and Fronsdal \cite{FF},
stating, in its original form in $D=4$, that the product of two
(scalar or spinor) singleton representations decomposes into an
infinite sum over all possible massless representations. This
result, which can be generalized to other values of the AdS dimensionality $D$, can be interpreted by
saying that any massless particle in AdS is a composite object made
of two singletons. The product of more than two singletons gives
massive representations. However, to the best of our knowledge, a
careful analysis of this decomposition has not been performed yet.

The scalar singleton is a very simple representation. To construct it, one starts from the commutation relations of the
$\mso(D-1,2)$ algebra, given by
\bea
\left[ M_{AB},M_{CD}\right]=i\eta_{BC}M_{AD} + 3~~ {\rm perm.}~,~~~~~~M_{AB}=-M_{BA}=(M_{AB})^\dagger ~,
\eea
where $A=0',0,1,...,D-1$ and $\eta_{AB}=\rm{diag}(-,-,+,...,+)$.
The maximal compact subalgebra $\mso(2) \oplus \mso(D-1)$ corresponds to time translations, generated by $E=M_{0' 0}$, and spatial rotations, generated by $J_{rs}=M_{rs}$) in AdS.
The remaining generators can be recast in the form of ladder operators $L^{\pm}_r=M_{0r}\mp i M_{0'r}$, raising or lowering the energy by one unit.
The algebra can be accordingly rewritten as follows
\bea
&&\left[E, L_r^{\pm}\right]=\pm  L_r^{\pm} ~~~;~~~ \left[J_{rs},L^{\pm}_t\right]= 2i\d_{t[s} L^{\pm}_{r]}\cr
&&\left[L^-_r, L^+_s\right]=2(iJ_{rs}+\d_{rs}E)\cr
&&\left[J_{rs},J_{tu}\right]=i\d_{st}J_{ru} + 3~~ {\rm perm.}
\label{algebra}
\eea
In general, a UIR of $\mso(D-1,2)$, denoted by $\mathcal D\left(e_0, {\bf j}\right)$ , is characterized by a lowest weight state $\vert e_0,{\bf j}\rangle$ carrying $\mso(D-1)$ spin {\bf j} and being annihilated by $L^-_r$, thus carrying minimal energy eigenvalue, denoted by $e_0$. Positivity of norms imposes bounds on the energy (for fixed spin). In the case of scalar representations, one starts from the lowest weight state obeying
\bea
(E-e_0)\vert e_0\rangle =0~~~;~~~
 J_{rs}\vert e_0\rangle=0~~~;~~~ L^-_r\vert e_0\rangle=0
  \eea
and build the generalized Verma module
\bea
\mathcal V\left(e_0,0\right)\equiv\left\{L^+_{r_1} \dots L^+_{r_n}\vert e_0\rangle\right\}_{n=0}^{\infty}\ .
\eea
Unitarity implies $e_0\ge \e_0\equiv \frac{D-3}{2}$ or $e_0=0$ (corresponding to the trivial representation). One can prove that, for the special value $\e_0=\frac{D-3}{2}$ saturating the unitarity bound,
\begin{itemize}
\item{
$L^+_s L^+_s \vert \e_0\rangle$ is itself a lowest weight state (it's a singular vector)}
\item{$ L^+_{r_1} \dots L^+_{r_n}L^+_s L^+_s\vert \e_0\rangle$ is normal to all states (it's a null vector)}
\end{itemize}
The scalar singleton is then the subspace obtained by factoring out the trace terms (the maximal ideal) from the generalized Verma module, \emph{viz.}
\bea
\mathcal D\left(\e_0,0\right)\equiv\left\{ L^+_{\{ r_1}\dots L^+_{r_n\}}\vert \e_0\rangle\right\}_{n=0}^{\infty}~~~,~~~{\tiny \e_0=\frac{D-3}{2}}~.
\eea
\vskip 10pt
In string theory, bound systems of singletonic particles arise as an effective description of rotating string and brane configurations in AdS.
\begin{center}
\begin{figure}
\begin{center}
\includegraphics[width=.7\textwidth]{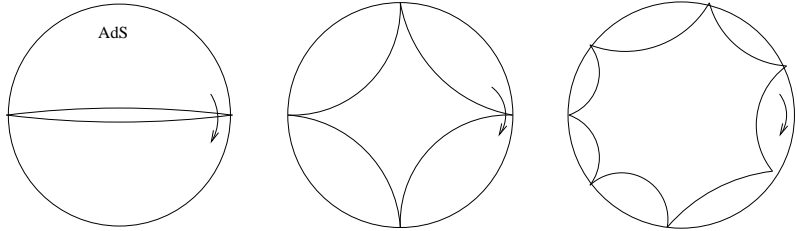}
\end{center}
\caption{2-cusp, 4-cusp and 8-cusp folded strings rotating in AdS.}
\label{fig:1}
\end{figure}
\end{center}
The simplest such solution is the folded rotating string with $E\sim S$, introduced by Gubser, Polyakov and Klebanov \cite{GPK}, stretching towards the boundary of AdS and forming cusps at the endpoints, where the classical energy and spin densities are concentrated. A semiclassical treatment of quantum fluctuations around this solution has been performed by Frolov and Tseytlin \cite{FT} and the generalization to the case with N-cusps has been discussed by Kruczenski \cite{Kruczenski}.
An important feature of these solutions is that they are effectively tensionless at the folded cusps.\\
In \cite{perjohan}, by revisiting the quantum fluctuation analysis, and paying particular attention to the behavior of the normal-coordinate fluctuation fields close to the cusps, two of the present authors have shown that the 2-cusp string admits an effective description as a bound state of two solitons, each corresponding to a wave-function localized to the cusps. The deviation from being (free) singletons is measured by the effective tension. Therefore, it is natural to expect that, in the tensionless limit, the system will be described by free singletonic particles living close to the boundary of AdS, \emph{i.e.} in a region of the bulk that asymptotes to the Dirac hypercone as seen by an observer at the center-of-mass of the rotating brane. This observation generalizes to the case with N cusps and also to rotating $p$-branes, that collapse in the directions transverse to motion and effectively behave as strings.

In \cite{perjohan}, by combining the above results with the old idea that energetic extended objects can be equivalently described by a set of partons, it was conjectured that the microscopic degrees of freedom of $p$-branes in AdS in the tensionless limit are singletonic partons. Moreover, reconsidering
the discretized $(0+1)$-dimensional Nambu-Goto action for a $p$-brane in AdS, taking into account the fact that isolated singletonic particles are only observable close the boundary of AdS, it was proposed that, in order to uncover the singletonic nature, the tensionless limit must be taken together with a zero radius limit in AdS. The geometric hypercone constraints then combine with the discretized version of the $p$-brane gauge symmetry generators into an $Sp(2N)$ gauge group, containing the conformal $Sp(2)$ symmetries of each parton as well as large $S_N$ transformation. The resulting system of discrete partons is then described by an $Sp(2N)$ gauged sigma model. One can think of its spectrum as a singleton gas.

A heuristic continuum-limit of the phase-space action describing N partons, discussed in \cite{perjohan}, suggests that, at least in the case $D=7$, processes involving an arbitrary number of partons could be described by
a gauged WZW model on a coset of the type
\bea
\frac{\widehat{\mso}(D-1,2)_{-\e_0}}{\widehat{\mathfrak{h}}_{-\e_0}}~,~~~~~ {\rm with}~ \e_0=(D-3)/2~,
\label{coset}
\eea
where the subalgebra $\mathfrak{h}$ should be chosen such that the gauged model contains only singletons and their composites.
In general $D$, we have proposed to gauge the maximal compact subalgebra \cite{ourpaper,us2}
\bea
\hat{\mathfrak{h}}={\widehat \mso}(D-1)_{-\e_0}\oplus {\widehat \mso}(2)_{-\e_0}~.
\label{cosetII}
\eea
This coset, with $D=5$, has also been considered for the description of the string worldsheet in AdS$_5$ in the ultrarelativistic limit \cite{mikhailov}.
The choice of the subcritical level in (\ref{coset}), (\ref{cosetII}) leads to a coset with vanishing central charge and singleton primary fields with vanishing conformal weight. Both properties are independent of the dimensionality $D$.

We would like to stress here that the level $k=-\e_0$ is fractional for even $D$.
Since the behavior of CFTs with fractional level is not well understood, apart from a few very simple examples with compact groups (i.e. $SU(2)_{-1/2}$ \cite{mathieu}), the analysis of this proposal is far from obvious.
\vskip 10pt
In our paper \cite{ourpaper} we have studied the $\widehat {\mso}(D-1,2)_{-\e_0}$ WZW model.
The main observation is that the special value $k=-\e_0$ is uniquely selected by the requirement that the symmetric traceless rank 2 tensor
  \bea
 V_{AB}(z)=\Big(M_A^{~C}(z)M_{BC}(z)\Big) - \frac{\eta_{AB}}{D+1}\Big(M_{CD}(z)M^{CD}(z)\Big)
 \eea
 is a WZW primary. The physical WZW primaries must decouple from $V_{AB}$,
 \bea
 \langle  \rm{phys'}\vert V_{AB}\vert \rm{phys}\rangle=0~ .
 \label{dec}
 \eea
 As explained in \cite{ourpaper}, this condition is the natural affine analogue of the $SO(D-1,2)$-covariant version of the equation of motion for the conformal particle, \emph{i.e.} $L^+_s L^+_s \vert \e_0\rangle\sim 0$. However, at the affine level, other solutions than the singleton are possible.
 In fact, one can prove that, in general $D$, the decoupling condition (\ref{dec})
is solved by infinitely many scalars
  \bea
 \vert e_0=P\e_0\rangle, ~~~~~~P=0,1,2,\dots
 \eea
 defined by the $P$-twisted lowest weight conditions
 \bea
 &&\left(L^-_r\right)_n\vert e_0\rangle=0, ~~~ n\ge -P ~~~;~~~   \left(L^+_r\right)_n\vert e_0\rangle=0,~~~ n\ge P\cr 
&& \left(J_{rs}\right)_n\vert e_0\rangle=0,~~~ E_n\vert e_0\rangle=\d_{n,0}~P\e_0\vert e_0\rangle, ~~~n\ge 0
 \eea
 In the case $P=0$ these conditions define the unity operator. For $P=1$, the twisted scalar can be easily interpreted as the singleton ground state, by noticing that, with respect to a standard affine extension, an extra singular state, $L^-_{r,-1}\vert \e_0\rangle$, has been modded out.
 In the case $P=2$, the $P$-twisted scalar is identified with the scalar ground state for the massless sector, while for $P>2$ cases we have massive ground states.
 In the case $P=2$, it is interesting to observe that a single ground state generates the whole set of massless representations and one can hop from one massless state to another with spin of one unit higher by applying the affine algebra operator $L^+_{r,1}$.
 This is in some sense analogous to what happens in a higher spin algebra (where all massless fields sit in the same multiplet), but in the more familiar setting of affine Kac-Moody (KM) algebras. It would be nice to investigate these aspects further to see whether an affine Lie-algebraic setting for HSGT is possible.\\
In $D=3,4$ one can show that the spinor singleton and its composites also solve the decoupling condition (\ref{dec}).

The presence of the whole set of $P$-twisted scalars in the physical spectrum of the $\widehat{\mso}(D-1,2)_{-\e_0}$ WZW model can be understood by considerations based on spectral flow.\\
The spectral flow is
 an invariance of the current algebra (the affine version of (\ref{algebra}))
\bea
&&[J_{rs,m},J_{tu,n}]=i(\d_{st} J_{ru,m+n}+ 3~{\rm perm.})+2km\d_{t[r}\d_{s]u}\d_{m+n,0}\cr
&&[E_m,E_n]=km\d_{m+n,0}\cr
&&[L^-_{r,m},L^+_{s,n}]=2(iJ_{rs,m+n}+\d_{rs}E_{m+n})-2km\d_{rs}\d_{m+n,0}\cr
&&[L^\pm_{r,m},L^\pm_{s,n}]=0\cr
&&[E_m,L^\pm_{r,n}]=\pm
L^\pm_{r,m+n}~~~;~~~[J_{rs,m},L^\pm_{t,n}]=2i\d_{t[s}L^{\pm}_{r],m+n}\cr
&&[J_{rs,m},E_n]=0
\eea
under the transformations
\bea
L^{\pm}_n\longrightarrow L^{\pm}_{n\mp w}~~~;~~~E_n\longrightarrow E_n+k w\d_{n,0}~~~;~~~J_{rs,n}\longrightarrow J_{rs,n}~.
\eea
One can easily see that the $P$-tupletons are all connected by the spectral flow.
For this reason, if the theory preserves spectral flow, all $P$-twisted scalars must be in the spectrum.
In \cite{ourpaper}, in the special case $D=4$, we have considered a free field realization of the WZW model in terms of symplectic bosons. This analysis allowed us to compute, for instance, the fusion rules and to check their compatibility with the ones conjectured by spectral flow considerations.\\
Another interesting observation related to the spectral flow is that the Goddard-Kent-Olive (GKO) conditions for the maximal compact gauging
\bea
E_m \vert\psi\rangle=0~~~,~~~J_{rs,m}\vert\psi\rangle=0~~~(m>0)
\eea
are invariant under spectral flow.
Therefore, maximal compact gauging would be preferred if the theory preserves spectral flow.
\vskip 10pt
In \cite{us2} we initiated the study of the spectrum of the coset
 \bea
\frac{\widehat{\mso}(D-1,2)_{-\e_0}}{\widehat{\mso}(D-1)_{-\e_0}\oplus \widehat {\mso}(2)_{-\e_0}}~,~~~~~ {\rm with}~ \e_0=(D-3)/2~,
\eea
focusing on the one-singleton sector.
By applying the GKO gauging procedure we found that at Virasoro level 1 all the states that survive the gauging are KM singular and decouple.
As a result, there are no level 1 excitations left and the spectrum at level 1 is empty.
However, at higher levels it is possible that nontrivial excitations, i.e. states that are gauge invariant but not singular, arise. This analysis is still work in progress \cite{us2}. Moreover, it would be nice to perform a BRST treatment of the gauging, instead of the GKO construction, since the two are not necessarily equivalent. An interesting open problem is also to reproduce the results obtained by this abstract analysis through a free field realization of the model (possible in $D=4,5,7$, even if in the last two cases one needs an additional gauging).
\vskip 10pt
To summarize,
 the maximal compact gauging of $\widehat{\mso}(D-1,2)_{-\e_0}$ seems to be a good candidate for describing the singletonic degrees of freedom of tensionless branes in AdS.
 The choice of the level $k=-\e_0$, set by lifting the singleton equation of motion to the affine case, is the key point in our construction.
 The ungauged $\widehat{\mso}(D-1,2)_{-\e_0}$ WZW model displays striking features, in particular the need for a
generalized definition of primary fields,
supported by the nontrivial action of the spectral flow.
The massless sector, with $P=2$, is of interest for HSGT.
All massless primary fields are generated by one LWS and it is possible to
 hop from one massless field to another by acting with $L^\pm_{r,1}$.
 This might point to a possible affine Lie algebraic setting for HSGT.
 In our paper \cite{ourpaper} we have also considered a free field realization of the $\widehat{\mso}(3,2)_{-1/2}$ WZW model in terms of symplectic bosons. In particular, we have computed the fusion rules and checked their consistency with the predictions obtained from spectral flow considerations.

{\bf Acknowledgements:}
  This work is supported in part by the European Community's Human Potential Programme under contracts MRTN-CT-2004-005104 `Constituents, fundamental forces and symmetries of the universe', MRTN-CT-2004-503369 `The Quest For Unification: Theory Confronts Experiment' and MRTN-CT-2004-512194 `Superstring Theory';  and by the INTAS contract
03-51-6346 `Strings, branes and higher-spin fields'.
  L.T. would like to thank the Foundation Boncompagni Ludovisi, n\'ee Bildt, for financial support during her two-year stay in Uppsala, where part of the results presented in this paper were obtained. Finally, the research of P.S. was supported in part by a
visiting professorship issued by Scuola Normale Superiore; by INFN;
by the MIUR-PRIN contract 2003-023852; and by the NATO grant PST.CLG.978785.

\end{document}